# SYSTEMS ENGINEERING FOR CIVIL TIMEKEEPING

Rob Seaman[*]

The future of Coordinated Universal Time has been a topic of energetic discussions for more than a dozen years. Different communities view the issue in different ways. Diametrically opposed visions exist for the range of appropriate solutions that should be entertained. Rather than an insoluble quandary, we suggest that well-known systems engineering best practices would provide a framework for reaching consensus. This starts with the coherent collection of project requirements.

**INTRODUCTION**

One can speculate on the origins of human nature and our capacity for creative problem solving. Presumably there are strong evolutionary pressures as represented in well-worn phrases like "look before you leap" and "necessity is the mother of invention". The most fundamental observation about how humans address complex challenges is perhaps that we almost inevitably seek to think about potential solutions before completely analyzing the problem in front of us. This instinctual cognitive short circuit may have had survival value when evading predators on the savanna, but serves us less well in the setting of a complex technological civilization.[1]

Civilization's infrastructure has another widespread characteristic shared with biological evolution. Each generation of technology is derived from the one before. Our clocks are sexagesimal because our parents' and grandparents' clocks were, all the way back to the Sumerian[†] concepts of time 5,000 years ago. The combination of these two effects motivates technological solutions that are highly stable yet exhibit sudden phase transitions as the prior paradigm may be replaced entirely from one generation to the next.

Finding a robust consensual solution to an esoteric engineering problem thus requires navigating between the extremes of instinctual snap judgments and inherited tradition. *Systems Engineering* is the discipline that provides tools for managing this challenge, to reduce the brittleness of the decision-making process and to promote solutions that multiple stakeholders with diverse requirements and worldviews can jointly accept. Systems Engineering does not seek the "best" solution, rather it seeks a family of satisfactory solutions whose benefits and shortcomings can be traded-off against one another using unbiased techniques balancing cost, schedule, performance and not least, the risks of each proposal.

---

[*] Senior Software Systems Engineer, National Optical Astronomy Observatory, 950 N Cherry Ave, Tucson, AZ 85719.
[†] http://www.nytimes.com/2010/11/23/science/23babylon.html

**REQUIREMENTS VERSUS SPECIFICATIONS**

One such esoteric engineering challenge is *civil timekeeping* – seemingly simple, but deceptively complex. Civil timekeeping provides another example of how humans grapple with problems. When contemplating an issue, we will often fixate on one small aspect of the situation and ascribe special importance to it. In the case of civil timekeeping this is the "leap second".

The problem of civil timekeeping is not leap seconds, rather the *problem* is how to meet a wealth of engineering requirements that includes aspects of atomic timekeeping and Earth orientation. The current *solution* to the problem of civil timekeeping happens to include the issuance of leap seconds. Ceasing leap seconds will not and cannot make the underlying engineering requirements go away. Requirements are just that – required. There are only three ways to handle any engineering requirement:

1. satisfy it,
2. issue an explicit lien against the requirement, or
3. reach consensus among the stakeholders that it is not, in fact, a requirement.

What is a requirement? It is not a specification. Requirements are inherent in the problem concept, whereas a specification describes some aspect of a proposed solution. A requirement is independent of human foibles and opinions (other than as humans themselves enter as part of the problem description). One **discovers** requirements. They exist independently of the observer. A requirement is also not a preference and is not something to be prioritized. Each solution that is proposed must address every requirement.

An example may help. This section is being written on an airplane. The problem concept may be stated as a flying vehicle. Two things among many that are required are lift and thrust. Wings, however, are not required. Wings are specified as a component of one class of solution, airplanes. But neither helicopters nor balloons have fixed wings. They do, however, develop lift. Jet engines are also not required. Helicopters can use their rotors to develop both lift and thrust, coupling those requirements in this class of solution, whereas hot-air balloons rely on the wind for their thrust. (A tethered balloon is not a vehicle, although its passengers may be said to fly.) That a dirigible blimp develops thrust reveals the separability of these two requirements in the more general lighter-than-air case. Lift is required else the vehicle could not be said to "fly", an element of the problem concept, but wings are only specified, not required. Note that this is not a value judgment – we passengers find wings very important nonetheless. But wings are important only because they provide a way for the vehicle to meet the requirements of air travel.

In the case of civil timekeeping, leap seconds have been specified as part of the currently codified and deployed solution.[2] What then is required? We will discuss this below.

**SYSTEMS ENGINEERING BEST PRACTICES**

The intent of this document is not to lecture on systems engineering. Rather, the goal is to highlight a few central concepts of systems engineering as they may apply to the discussion of the future of Coordinated Universal Time. In particular, it is to advocate for a more coherent community-wide examination of this issue and the implications of the different outcomes.

What is *systems engineering*? It is one organized method for reaching decisions when presented with complex challenges, convoluted environmental constraints, and diverse stakeholders. Other methods include such things as "Creative Problem Solving,"[3] "Principled Negotiation,"[4] and in a narrower context, software development processes (for one among many examples).[5]

More than a decision-making protocol, systems engineering works with project management techniques to actually accomplish the fruits of those decisions.

**Systems Engineering Management Plans**

Systems engineering is a type of planning discipline. In a formal exercise of these techniques that plan is always written down. The document format varies widely, but maps to the structure of the particular process being followed (Figure 1).

|        | Name | Order in which the documents are started | Order in which the documents are finished |
|--------|------|----|----|
| Doc. 1 | Problem Situation | 1 | 8 |
| Doc. 2 | Customer Requirements | 5 | 2 |
| Doc. 3 | Derived Requirements | 6 | 4 |
| Doc. 4 | System Validation | 7 | 5 |
| Doc. 5 | Concept Exploration | 3 | 3 |
| Doc. 6 | Use Case Model | 4 | 1 |
| Doc. 7 | Design Model | 8 | 6 |
| Doc. 8 | Mappings and Management (schedules) | 2 | 7 |

**Figure 1. A Systems Engineering Management Plan (SEMP)[6] comprises multiple documents (sections) and is intended for diverse stakeholders.**

In this process it is explicitly recognized that the order in which the documents are read differs from the order in which they were written, and in fact that the documents are started in one order but finished in another. One can quibble over the details of any formal process, but rather consider some of the general features expressed in this table. The first document to be started is a description of the *Problem Situation*, which is to say an expression of the underlying nature of the system in question. Perhaps surprisingly, this is the last document to be finished. The idea here is that throughout the engineering process – whatever process – the understanding of the problem space is extended and deepened. The Problem Situation is also the first document to be read. Whether one is conceptualizing and planning a system or rather investing in understanding someone else's plans, the place to start is with a description of the problem space.

At the far end of the writing process lies the *Design Model*, that is, an expression of the solution that has been refined and selected as proposed by the plan. This is also near the end of the plan documents as read by stakeholders – only the project schedules, budgets and other supporting documents follow, perhaps as an appendix. In short, the structure of the document formalism logically separates the characterization of the problem from its proposed solution(s).

In between the statement of the problem and the proposed design for a solution lie a variety of documents and activities that capture the conceptualizing invested in the engineering process. In the example of the specific process illustrated in Figure 1, these include descriptions of the *Customer* and *Derived Requirements* (sometimes called the *non-functional* and *functional requirements*) and *Use Cases* that capture the desired behavior of the system. *System Validation* is performed – does the system satisfy the requirements? Depending on the complexity and breadth of the problem being addressed, these documents may be quite extensive and detailed, in particular the *Concept Exploration* activities may involve significant research. But in even the simplest system the questions posed remain the same:

1. What is the problem at hand?
2. How must the system behave?
3. What do the stakeholders require?
4. How will compliance be ensured?

**All Engineering Processes Share Fundamental Characteristics**

Figure 1 expresses just one of many possible systems engineering processes and formal planning techniques. Perhaps there are others that proceed in a different order and comprise different sorts of documents and underlying activities? Rather, a broad survey of systems engineering practitioners has demonstrated that all widely used such processes are generally equivalent. The mnemonic that captures this notion is *SIMILAR*,[7] which maps well onto many other system development processes:

- **S**tate the problem
- **I**nvestigate alternatives
- **M**odel the system
- **I**ntegrate interfaces
- **L**aunch the project
- **A**ssess performance
- **R**e-evaluate your solution

For the purposes of this paper, we need not delve too deeply into the details of this equivalence, but note again the clear separation between 1) effort invested in understanding the problem space, and 2) work to design a solution that is compliant with the engineering requirements so discovered. This mnemonic captures another key aspect (in "Re-evaluate your solution") of systems engineering best practices, namely that engineering is an iterative exercise during all phases from design and development to operations and maintenance. In particular, an assertion that a particular solution will remain viable for all time is rarely appropriate. Operating on this assumption would guarantee widespread future disagreements and the frequent revisiting of supposedly settled civil timekeeping issues.

**REQUIREMENTS OF CIVIL TIMEKEEPING**

How can these techniques be applied to the problem of civil timekeeping? This brief paper cannot hope to capture the richness of an appropriately detailed Systems Engineering Management Plan. However, perhaps a bit of clarity can be achieved in the stating of underlying concepts to inform future discussions.

Consider again Figure 1, in particular, the order in which the planning documents (or sections or chapters) are finished. It was previously pointed out that the final document finished, the *Problem Situation,* is also the first one started, but what are the **first** documents finished? In this formalism it is the *Use Case Model* and the *Customer Requirements*, that is: What is the consensus among stakeholders for how the resulting system must behave and for how people will interact with it? In the case of the frequently zealous civil timekeeping discussions over the past several years, this is precisely the initial phase of the discussion that has been omitted. By focusing exclusively on proposed partisan solutions, opportunities for a durable consensus were missed. Time was lost that could have been better spent in engaging with stakeholders and in characteriz-ing the common problem. It may even be that some variation of one of the proposed solutions could satisfy the engineering requirements—but how would this ever be discovered if the requirements are never written down, and if the stakeholders are never consulted for comment?

No solution will be proposed in this paper. The author is confident that a consensus solution can be found that is appropriate to our time and for our purposes. Instead of solutions, several requirements on the problem of civil timekeeping will be asserted. Indeed, a simple statement of faith in the existence of an appropriate solution to the problem leads immediately to discovering the first such engineering requirement:

> *A. Forward as well as backward compatibility must be preserved.*

Numerous discussions have pointed out the long-term quadratic behavior of the tidal slowing of the Earth's rotation.[8] No matter what mechanism is implemented now or in the future this will always be a characteristic behavior of civil timekeeping on Earth. The preservation of continuity over transitions from one civil timekeeping standard to the next is a clear engineering requirement. Remember, a requirement must either be satisfied – or an engineering *lien* must be placed explicitly on the requirement. A lien does not mean that the requirement can be ignored, rather the opposite. A lien means that an explicit mitigation plan exists to deal with the implications of not having met the requirement. Proposed solutions that are brittle over the long term are significantly disadvantaged by this requirement. In fact, implementing a lien is often more onerous and expensive than meeting the requirement in the first place, especially over the long term.

**Engineering Vocabulary for Civil Timekeeping**

The debate pertaining to a possible redefinition of Coordinated Universal Time (UTC) has been ongoing for more than a dozen years.[*] Unfortunately, these discussions have been less successful than they might have been. One reason is a lack of outreach – consider that the ITU-R has only held one public meeting on this topic the *Colloquium on the UTC Time Scale*[†] in Torino Italy, 28-29 May 2003. A more fundamental reason is a lack of shared vocabulary for constructing the discussions themselves. This needed vocabulary includes highly technical jargon, but starts with the most basic building blocks of the debate.

As this discussion continues, the language of systems engineering can make the relationships between frequently bandied terms clearer:

- *Problem statement*: civil timekeeping
- *Proposed solution*: UTC
- *Feature of solution*: intercalary steps

That is, we are asserting that timekeeping for civilian purposes comprises a single coherent problem space. Designing and deploying one-or-more timekeeping standards and systems will provide a solution to this problem. The current solution to this problem is Coordinated Universal Time. UTC meets the system requirements via intercalary steps,[‡] currently in the form of leap seconds (whether these are actual leaps or merely a representational overlay on TAI).

---

[*] http://groups.google.com/group/sci.astro.fits/msg/9326ad192333a560
[†] http://ucolick.org/~sla/leapsecs/torino/
[‡] Are intercalary steps *required*? No, they are an emergent property of the problem space. The current UTC standard happens to specify leap seconds, but if it were to be redefined to embargo these, the intercalary steps would continue to accumulate until some other accommodation is made to make the equivalent adjustment. It is the attempt to ignore this

Two key vocabulary words may be understood through examples:

- *Requirement*: civil time approximates mean solar time (see below),
- as opposed to a *Specification*: leap seconds.

Systems engineering terminology can help to clarify talking points and avoid confusion and delay. This confusion started at the very beginning when the issue was characterized as the "problem with leap seconds." As has been explained, leap seconds are not the problem in the engineering sense. It goes further than this, however: the problem is not even UTC. Rather, the problem is *civil timekeeping* itself – that is, the worldwide system of timekeeping suitable for civilian purposes.

By focusing on leap seconds and UTC, the actual engineering requirements of civil timekeeping are obscured and become confused with requirements for other types of timekeeping. This is a second engineering requirement:

  B. *There is more than one kind of time and thus of timekeeping.*

Note an immediate benefit from clearly stating the problem and its requirements; namely, further discussions can steer clear of things that are not part of this particular type of timekeeping. Clocks come in many forms, but some kinds are more directly pertinent to civilian use than others.

**The Stakeholders are Many**

While the author strongly recommends systems-engineering best practices as a solid basis, readers serve themselves well by asking these three questions, whatever their decision-making process:

- What is civil timekeeping?
- What is required to implement a worldwide civil timekeeping system?
- Who are the users of civil timekeeping?

Let us focus now on the third question. The short answer is that there are seven billion civil timekeeping users. Focusing disproportionately on leap seconds serves to artificially narrow the list of stakeholders. Teasing this one thread out of its proper broader engineering context leads to the omission of affected communities, rather than their inclusion. User exclusivity simply does not arise when the true breadth of the problem space and its stakeholders is realized, leading to the next requirement:

  C. *Timekeeping users are ubiquitous.*

There is certainly not room at the table for seven billion seats. But there should be room for representatives from many more communities than currently consulted by the narrow interests of the Radiocommunication Sector of the International Telecommunication Union.[9]

This requirement warrants a bit more discussion. The short answer was that there are seven billion stakeholders in civil timekeeping. The longer answer is that very few will have enough familiarity with timekeeping issues to participate in detailed discussions. This is where systems engineering techniques truly shine. A formal but lightweight engineering process is particularly

---

requirement that results in its being reasserted. If an alternate mitigation strategy is sought that has no intercalary steps, then the requirements must be faced head-on.

useful for structuring the gathering of requirements from diverse stakeholders with varying levels of knowledge and expertise. In fact, the alternative to using such techniques is precisely to exclude such stakeholders from participating in the decision-making process at all.

**The Use Cases are Diverse**

Civil timekeeping cannot be all things to all people, yet it must serve many purposes. The central problem with failing to follow a coherent process for collecting engineering requirements is that communication about the issues turns into an exercise in imagination. One party or another will indicate that it knows what is "best" for all purposes and all users. By treating requirements like preferences, some narrow aspect of the problem will be asserted to have a "higher priority" than all others. The reality is that the diverse and numerous requirements must all be satisfied at the same time.

A pertinent quote is attributed to Albert Einstein: "Make everything as simple as possible, but not simpler." To force a particular solution is to try to make things simpler than they really are. In the case of civil timekeeping, sometimes clock readings are needed to measure time intervals, and sometimes clock readings are needed to indicate time of day. Both aspects are inherent in the many use cases that pertain to civil time.

What are *use cases*? They are a broad selection of necessary behaviors that capture a sufficiently complex picture of how the system behaves across the community of stakeholders. This doesn't mean that all possible usages of clocks need be captured or should be given equal weight if that were even possible. But it does mean that any overly narrow list of clock functions will fail to meet the challenge.

Paraphrasing Einstein, the requirement on the complexity of timekeeping is:

> D. *Make clocks as simple as possible, but not simpler.*

In modern engineering processes it is predominantly use cases that are used to drive the discovery of requirements. A broad universe of civil timekeeping behaviors will inform a lengthy, but by no means vast, list of use cases. These use cases will include technical, scientific, cultural, legal, religious, philosophical, historical, logistical, economic and other aspects. With a system as widespread and fundamental as civil timekeeping, the list is indeed long. Similarities will become evident among diverse classes of usage. The major underlying engineering requirements will be relatively few in number—this paper makes a start.

**We Live on a Planet Lit by the Sun**

Some facts are so well known that they become invisible; in that case we may say that we "can't see the forest for the trees." On planet Earth nothing is more blatant than the Sun in the sky.[*] We know it is there even on a cloudy day. Somewhat more shy is the Moon, but we could infer its existence from the tides even without the ability to make a single celestial observation. It is unsurprising that the star that we orbit and the satellite that orbits us both affect the nature of timekeeping on Earth.

---

[*] The presentation corresponding to this paper at the colloquium *Decoupling Civil Timekeeping from Earth Rotation* included comparisons between timekeeping on the surface of the Earth and on Edgar Rice Burroughs' worlds of Pellucidar (at the Earth's core) and Barsoom (Mars). The issue is not science fiction – the issue is how humans would behave under those circumstances.

The alternative is to suggest that our clocks could be set to any hour of the day whatsoever. In this case the phrase "hour of the day" loses all meaning and the date itself becomes wholly arbitrary, rather than something delimited by intervening periods of nighttime. Readers might ask themselves why, exactly, the SI-second was chosen not only to have the same *name* as the second of sexagesimal time and angle representations – but also was specified to *approximate* the fraction of $1/86,400^{th}$ part of a mean solar day. If the SI-second had been chosen even so slightly different, as to be the $1/86,399^{th}$ part or the $1/86,401^{th}$ part of a day, there would have to be a leap second every day or a leap hour every decade.[*]

The question to ask here is not "How would we solve this?" The first question to ask is "What is the underlying engineering requirement?" That requirement is:

> E.  *Civil time must closely approximate mean solar time.*

Understanding what it means to say this is a requirement is the heart of the exercise in reaching consensus on how to move forward with civil timekeeping. A more accurate statement is: "Civil time *is* mean solar time", because this is really just a definition of terms. However, the notion of approximation is inherent in all engineering trade-offs. Remember, systems engineering does not seek the *best* solution, it seeks a family of *satisfactory* solutions whose cost, schedule, performance and risks can be traded off against each other. Trading off, however, does not mean discarding requirements, but rather a rebalancing of a proposed solution's accommodation of the requirements.

**Technology Evolves Rapidly**

By focusing on putative solutions rather than actual engineering requirements our options narrow to overemphasize currently perceived hurdles, especially of technology. For instance, the jargon word "POSIX"[†] is very rare in everyday speech, but very frequently heard on the so-called LEAPSECS[‡] mailing list (so-called because its name would more accurately reference UTC, rather than the narrow issue of leap seconds). Timekeeping standards are – or at least, should be – much longer lived. Greenwich Mean Time[10] (GMT) extends back to the Nineteenth Century[11] before computers and airplanes, automobiles or artificial satellites. Coordinated Universal Time[12, 13] (UTC) as an approximation to GMT[§] has been a standard for four decades, predating the Global Positioning System (GPS) and smart phones. The corresponding engineering requirement might be stated as:

> F.  *Civil timekeeping must support technologies not yet invented.*

A corollary to this is that we should seek instances of long-lived technologies as especially valuable exemplars when considering similarly long-lived standards. In the case of civil timekeeping one such technology is precisely that of the astronomical telescope, which has recently celebrated its $400^{th}$ anniversary. One should view the undisputed fact that redefining UTC would require rewriting large amounts of astronomical software as the "canary in the coal mine" for civil timekeeping.[14]

---

[*] By contrast, if the SI-second had been selected to be a value completely disjoint from this fraction of the mean solar day, the underlying laws of physics would retain exactly the same form. The overriding engineering requirement is supplied by those facts of nature that cannot be easily modified (mean solar time), not by those things that can (physics works out the same in the end).
[†] http://standards.ieee.org/develop/wg/POSIX.html
[‡] http://six.pairlist.net/mailman/listinfo/leapsecs/, messages from 2000-2007 at http://www.ucolick.org/~sla/navyls/
[§] Per CCIR Recommendation 460-4, "GMT may be regarded as the general equivalent of UT."

**History is long**

Astronomy is an ancient pursuit, but so are clocks and calendars.[15, 16] Any proposal to redefine the practice of civil timekeeping should be informed by related historical events. One particularly apt analogue was the Gregorian calendar reform.[17] Programmers will attest that calendar-related bugs continue to turn up in their software. This is not an argument to eliminate the Gregorian calendar as a common standard spanning all the world's cultures from centuries past to centuries yet-to-come. History does extend into the future.[18] Current timekeeping deliberations should take special care to properly value long-term requirements. This leads us to the final requirement to be asserted here:

> G. *We cannot un-ring the bell.*

Another way to say this is the Hippocratic dictum "*First, do no harm.*" A re-engineering effort that properly acknowledges the underlying system requirements retains the freedom to evolve further in the future. By choosing a new solution that meets the same coherent set of requirements as the original solution, a lightweight transition is effected that would also permit transitioning back to the *status quo* as necessary. Since the same underlying conceptual model is implemented, it preserves a "paper trail" for interpreting dates and times across the transitions.

On the other hand, if an abrupt transition is made to a completely different conceptual model then the signature of this discontinuity will be preserved forever. The interpretation of dates and times will require software and hardware that encapsulates both the old (*UTC is time-of-day*) and new (*UTC is atomic time*) rules. Since the new model is overly simplistic, any future transitions (guaranteed to occur by the ever-accruing embargoed leap seconds) will involve a similarly large **conceptual leap**. Which is to say that a proper systems engineering plan includes a discussion of risks and how these might be mitigated.

**WHEN IS A SOLUTION "GOOD ENOUGH"?**

This paper has served not only as a brief introduction to high-level systems engineering concepts, but also as an exercise in requirements discovery for civil timekeeping. The question naturally arises of what to do with the requirements so discovered:

> A. Forward as well as backward compatibility must be preserved.
> B. There is more than one kind of time and thus of timekeeping.
> C. Timekeeping users are ubiquitous.
> D. Make clocks as simple as possible, but not simpler.
> E. Civil time must approximate mean solar time.
> F. Civil timekeeping must support technologies not yet invented.
> G. We cannot un-ring the bell.

First note again that **requirements are not specifications**. Release any notion of a one-to-one correspondence. Requirement E is not an attempt to handcuff the process to "require" leap seconds. Leap seconds are not and cannot be required – they have rather been specified for several decades as an international standard. But other features of other possible solutions could serve to satisfy this requirement.

The existence of a requirement, however, ensures that a proposal to simply eliminate leap seconds will cause other behaviors to pop up somewhere else. Requirements exist separately from our comprehension of them – they are inherent in the problem space, in the behavior of the universe. Even if we don't recognize a particular requirement, it will exist to pester human systems.

Suppressing future leap seconds does indeed cause widely recognized issues. The embargoed leap seconds continue to accrue. At some future date, a larger intercalary adjustment will become necessary. This is the behavior of an engineering requirement spurned.

Well, okay – some may think – we will deal with the larger adjustment at some future date. The issue here is that a problem like civil timekeeping is characterized by many requirements simultaneously. An attempt to slay one requirement at a time is like playing *Whac-a-Mole*.[*] As each mole is pounded down another jumps up – in fact, more than one jumps up. In the case of civil timekeeping, when Requirement E is pushed down, not only does Requirement A jump up, but also F, and E itself refuses to stay down. Suddenly one is left looking three moles in the eye.

Which is to say that relaxing an engineering tolerance is one thing—and is a normal activity for a trade-off study. However, increasing a tolerance by three or four orders of magnitude is not relaxing it, but rather eliminating it. The precise requirement (to "approximate") one tries to eliminate reasserts itself because it is inherently required by the problem situation itself. And issues of continuity arise (Requirement A) as well as brittleness toward future technologies (Requirement F) and undoubtedly other requirements not listed here.

It is not so easy to completely redefine a problem space as to avoid all the pitfalls:

A. Sweeping issues under the rug of how the embargoed leap seconds will be accommodated in the future does not mitigate our responsibility to plan for this inevitable result—and it does not address short-term technical issues brought on by the decoupling of civil timekeeping from Earth rotation.

B. Precision time-interval and time-of-day are not the same things.

C. Focusing narrowly on high-technology users ignores a vastly larger population of more general users—especially considering that eliminating leap seconds does not address all high-technology use cases.

D. An oversimplified replacement standard does not address the inherent complexity of the underlying problem space.

E. Selective consideration of use cases that accommodate hour-long excursions from mean solar time does not address those use cases for which closer approximations are needed.

F. Paying too much attention to the limitations of current technologies and projects risks falling short in recognizing future trends and accommodating new opportunities.

G. Time and date standards are extremely long-lived and their impacts are broadly exposed to the lay public. A proposal that would redefine the essence of what timekeeping means must meet a very high level of due diligence.

**CONCLUSION**

Individual readers may disagree with the author's list of requirements. In fact, this is a central point of the exercise. By focusing on discussing use cases and requirements, consensus can be reached before any solution is proposed whatsoever. These requirements, and any others that may suggest themselves, should be debated. We should construct a coherent conceptual model for the problem space. We should seek solutions that satisfy all aspects of that model – or that dispose of shortcomings explicitly through liens. We should engage with all the stakeholders – certainly not just including the ones from our own communities. Building consensus and understanding of the problem space is not only the best way to conduct an exercise in systems engineering – it is the fastest way to identify and validate possible solutions.

---

[*] http://en.wikipedia.org/wiki/File:Whackamole.jpg